\documentclass[]{elsarticle}

\usepackage{lineno,hyperref}
\usepackage{multirow}
\usepackage{color}

\newcommand{\xin}[1]{\textcolor[rgb]{0,0,0}{#1}}

\newcommand{\wangql}[1]{\textcolor[rgb]{0,0,0}{#1}}

\begin{document}

\begin{frontmatter}

\title{Dual Windows Are Significant: Learning from Mediastinal Window and Focusing on Lung Window}



\author[cqu]{Qiuli Wang\corref{eqaulcon}}
\address[cqu]{School of Big Data and Software Engineering, Chongqing University}
\ead{wangqiuli@cqu.edu.cn}
\author[sjtu]{Xin Tan\corref{eqaulcon}}
\address[sjtu]{Department of Computer Science and Engineering, Shanghai Jiao Tong University}
\cortext[eqaulcon]{indicates equal contribution.}

\author[amu]{Chen Liu}
\address[amu]{Department of Radiology, Third Military Medical University (Arm Medical University)}

\begin{abstract}

    Since the pandemic of COVID-19, several deep learning methods were proposed to analyze the chest Computed Tomography (CT) for diagnosis.
    In the current situation, the disease course classification is significant for medical personnel to decide the treatment.
    Most previous deep-learning-based methods extract features observed from the lung window.
    However, it has been proved that some appearances related to diagnosis can be observed better from the mediastinal window rather than the lung window, e.g., the pulmonary consolidation happens more in severe symptoms.
    In this paper, we propose a novel Dual Window RCNN Network (DWRNet), which mainly learns the distinctive features from the \wangql{successive} mediastinal window.
    Regarding the features extracted from the lung window, we introduce the \emph{Lung Window Attention Block} (LWA Block) to pay additional attention to them for enhancing the mediastinal-window features.
    Moreover, instead of picking up specific slices from the whole CT slices, we use a Recurrent CNN and analyze successive slices as videos.
    Experimental results show that the fused and representative features improve the predictions of disease course by reaching the accuracy of 90.57\%, against the baseline with an accuracy of 84.86\%.
    Ablation studies demonstrate that combined dual window features are more efficient than lung-window features alone, while paying attention to lung-window features can improve the model's stability.

\end{abstract}


\end{frontmatter}


\section{Introduction}

COVID-19, which is resulted from the novel coronavirus, has been out-breaking, and the number of infected persons is reaching a new peak everyday \cite{article}. According to the World Health Organization (WHO) report, there are 41,104,946 confirmed cases and 1,128,325 confirmed deaths by the end of October 23, 2020~\cite{who-2020cases}.
The rapid spread of COVID-19 undoubtedly has long-range effects on the world and puts tremendous pressure on the current medical systems of every country. To relieve the docotors' workloads and speed up the treatment, the fast diagnose of COVID-19 and the decisions of its courses are strongly required.

Since the development of deep neural networks and the accumulation of lung Computed Tomography (CT) images infected by COVID-19, many CNN-based models are proposed to diagnose the COVID-19 automatically, such as \cite{9090149,2020Prior,WANG2020}. 
Although these methods have achieved some success to a certain extent for community-acquired pneumonia, few studies have pay attention to the disease course prediction, which plays a crucial role in CAD systems for COVID-19.
The difficulty of disease course prediction is that, as the successive step of COVID-19 diagnosis, it has to deal with the cases which have obvious symptoms caused by COVID-19.

\begin{figure}[htbp]
    \centerline{\includegraphics[width=100mm]{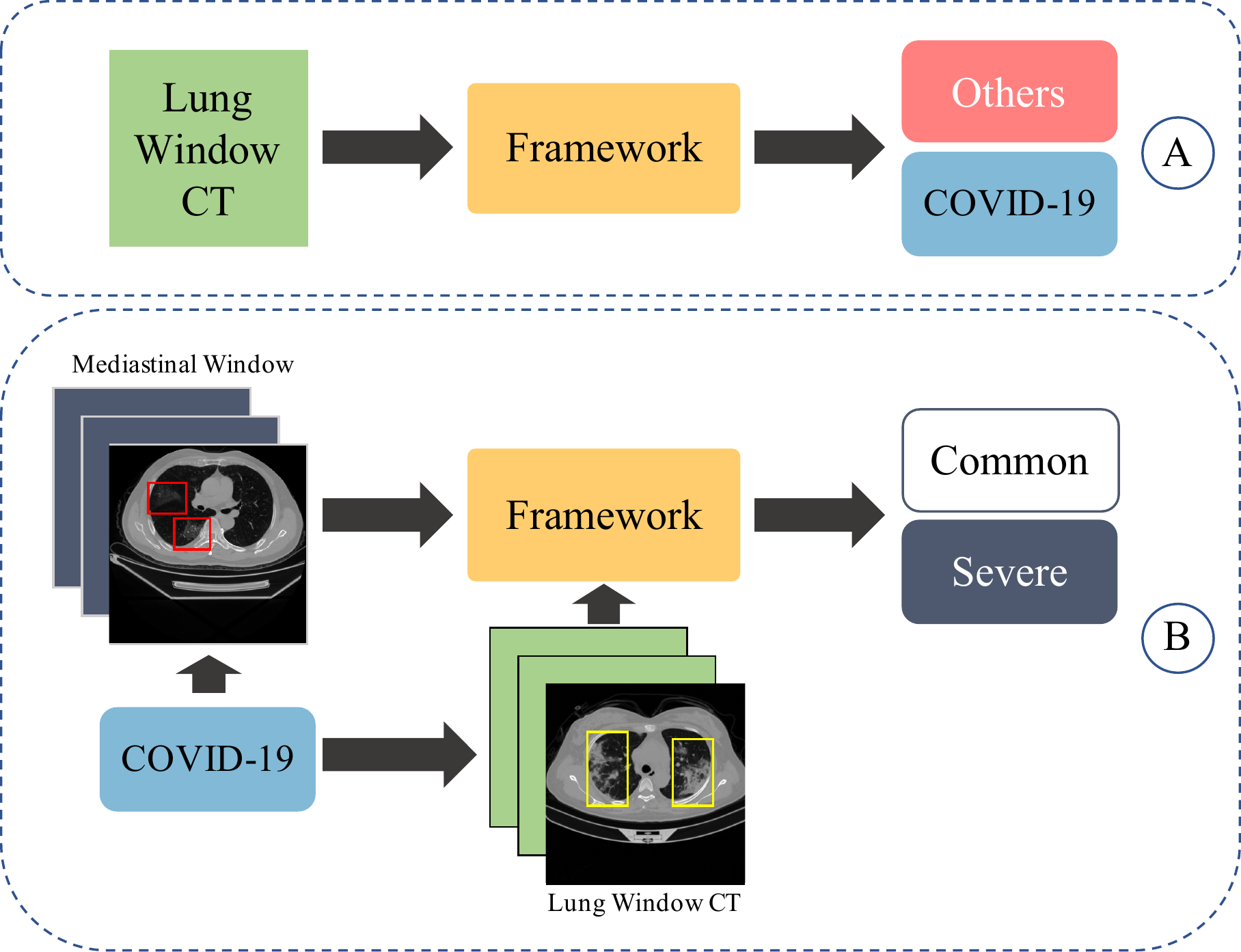}}
    \vspace{-0cm}
    \caption{A: Conventional methods learn lung window CT directly. B: DWRNet uses the mediastinal window as the main image window, and adds lung window features with Lung Window Attention Block.
    }
    \vspace{-0cm}
    \label{overflow}
    \end{figure}

\begin{figure}[htbp]
    \centerline{\includegraphics[width=140mm]{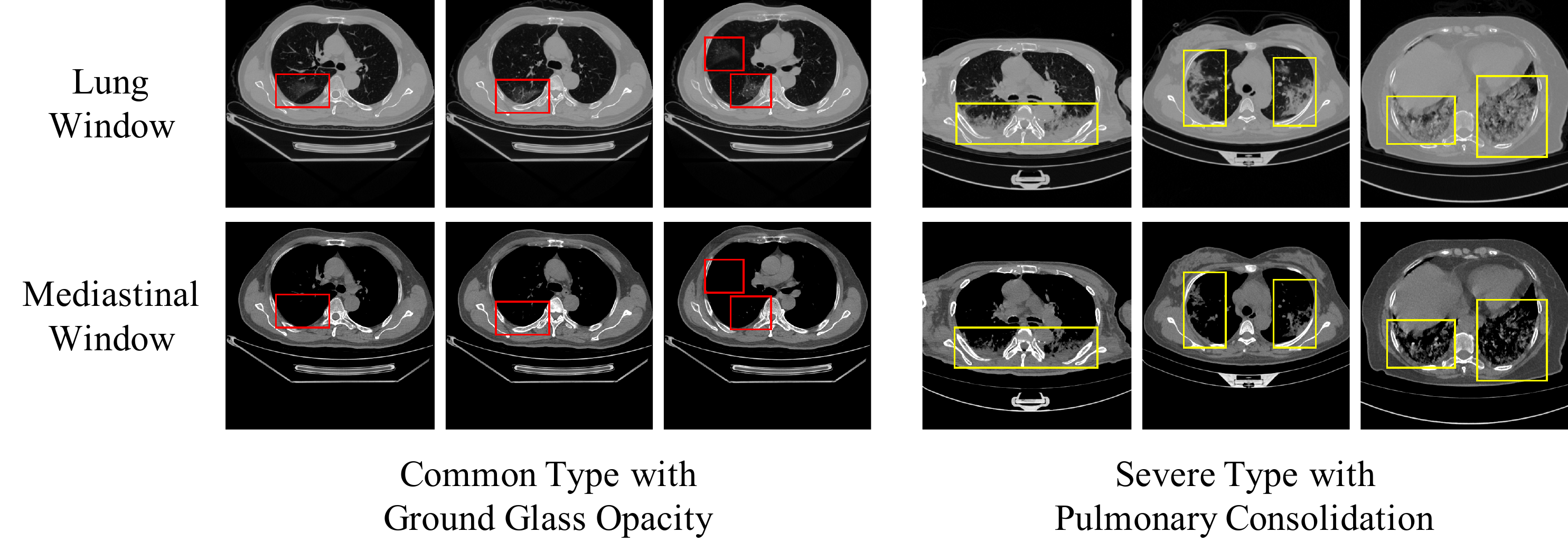}}
    \vspace{-0cm}
    \caption{Comparison between the lung window and mediastinal window. We can see that the common cases with GGO have apparent visual features with the lung window but are difficult to be diagnosed with the mediastinal window (red blocks). The severe cases show severe lung pathological changes when using the lung window. When changed to the mediastinal window, pulmonary consolidation can be clearly seen (yellow blocks).
    }
    \vspace{-0cm}
    \label{compare}
    \end{figure}

Conventional methods usually analyze CT slices with single-channel window, as shown in Figure~\ref{overflow}.A. However, based on the official \emph{Diagnosis and Treatment Protocol for COVID-19 Patients} \cite{china-protocol} provided by the China National Health Commission, Ground-Glass Opacity (GGO) is commonly observed in the common cases of COVID-19, and consolidation may occur in the chest of severe cases. 
Generally speaking, the lung window is the appropriate CT window for the observation of GGO \cite{macmahon2017guidelines}. Meanwhile, the consolidations can be better observed with the mediastinal window, which is a more stable method in the aspect of the inter-reader agreement \cite{2018Effect}, and show better performance in measuring the solid tissues \cite{2018Radiologic,Yao2016Value}. The effect of CT image windows is shown in Figure~\ref{compare}.
It is obvious that GGO can be seen with a lung window setting but disappears when using the mediastinal window setting. However, with the mediastinal window setting, solid pulmonary consolidation can be clearly seen and avoid the influences caused by other symptoms.

A straightforward solution to solve the above issue is to add another stream to learn useful features from the mediastinal window and combine two kinds of features, \emph{i.e.}, lung-window features, and mediastinal-window features. 
However, there are high chances that both kinds of features have some repeated information since they are just extracted from the same CT HU values. 

In this paper, we propose a Dual Window RCNN Network (DWRNet) based on mediastinal window image. It also add additional lung window features to the framework by attention mechanism. The overflow of DWRNet is shown in Figure~\ref{overflow}.B.
Previous \xin{works} \wangql{learn} most information from the lung window. \wangql{Unlike these stuides,} our framework learns the main features from the mediastinal window, then \xin{just} uses lung window features as the attention map by designing a Lung Window Attention Block (LWA Block).
\wangql{Moreover, instead of picking up specific CT images, the proposed DWRNet analyze two successive CT windows as videos and learn the discriminative features from the whole slices.}
Experiments show that using the mediastinal as the main CT window can significantly improve the overall performances, and additional attention map from the lung window makes the framework have better performances and more stable.

In summary, this paper has following contributions:
\begin{itemize}
    \item To our best knowledge, this is the first work to emphasize the importance of features observed from mediastinal window for disease course diagnosis of COVID-19, which is able to obviously improve the diagnosis accuracy.
    \item We propose a dual window RCNN network to predict the courses of COVID-19 by dual windows. Dual window features are adaptively fused by the Lung Window Attention Block, and treated as successive video frames, which reduce the information redundancy.
    \item Experimental results show that the proposed model achieves high accuracy in courses prediction. Moreover, it demonstrates that learning from the mediastinal window and focusing on lung window is an ideal way to reach competitive results, meanwhile improve stability.
\end{itemize}

This paper is organized as follows:
In Section \ref{relatedwork}, we introduce some studies about this issue;
In Section \ref{methods}, we introduce the architectures of the framework;
In Section ~\ref{experiments}, we show the performances of the proposed framework, and analyze the experimental results;
We also discuss some facts in Section \ref{discussion} and draw the conclusions in Section \ref{conclusions}.

\section{Related Works}
\label{relatedwork}
\subsection{Computer-Aid Diagnosis Systems for Pneumonia}
Computed Tomography (CT) has been widely used for clinical pneumonia diagnosis \cite{Upchurch2017Community,XIE2018102}.
Unlike X-Ray images, CT can provide more details of lung tissue density, and avoid the shallows caused by ribs or other tissues.
In 2016, Shin \emph{et al.} \cite{shin2016deep} exploited three important, but previously understudied factors of employing deep convolutional neural networks to computer-aided detection problems: CNN architectures, dataset characteristics, and transfer learning. Later in 2018, Gao \emph{et al.} \cite{gao2018holistic} presented a method to classify interstitial lung disease imaging patterns on CT images and introduced normal lung rescale, high attenuation rescale, and low attenuation rescale. In 2020, the study in \cite{9078812} further fused demographic information with multi-window CT images. In a word, deep learning methods have been widely adopted for the analysis of chest CT.

Recently, COVID-19 has been a global pandemic and threatening all over the world. Chest CT, especially high-resolution CT, has been recommended as a major clinical diagnosis tool in the hard-hit region such as Hubei, China \cite{zu2020coronavirus}.
In 2020, Kang \emph{et al.} \cite{9086482} proposed a unified latent representation to explore multiple features describing CT images from different views fully for COVID-19.
Fan \emph{et al.} \cite{9098956} proposed a novel COVID-19 Lung Infection Segmentation Deep Network (Inf-Net ) automatically identify infected regions from chest CT slices.
\wangql{Xinggang Wang \emph{et al.} \cite{9097297} further proposed a weakly-supervised framework based on 3D CT volumes.}
However, existing methods seldom analyze the effect of different image window settings on deep learning methods for COVID-19. 
There are two main difficulties for \xin{CT-based} frameworks: First, some abnormal regions are not salient enough in the early stage, which may be ignored. Second, in the severe stage, the abnormal regions may look similar to some community-acquired pneumonia, which also results in some wrong diagnoses.

Invoked by the studies \cite{gao2018holistic}, we design a framework that can make the most of lung and mediastinal windows for the diagnosis of the COVID-19 disease course.
In our study, we use the mediastinal window as the main observation window and use the lung window as the enhancement for detecting severe symptoms such as pulmonary consolidation.
Generally speaking, two image window inputs usually require two branches of CNN, such as study \cite{simonyan2014two-stream}.
In our study, \wangql{two image windows are treated as video frames and analyzed by a 2D ResNet. Both CT windows share the same pre-trained ResNet-50 \cite{he2016deep}} as a feature extractor so that we can significantly reduce the burden of calculation.

\subsection{Attention Mechanisms}
Attention mechanisms help guide the networks to focus on some important and discriminative features to improve the performances for multiple computer vision tasks, such as image captioning \cite{Guo_2020_CVPR,you2016image}, semantic segmentation \cite{Chen2017SCA,Yu2018Learning,Choi_2020_CVPR}, crowd counting \cite{Jiang_2020_CVPR, sindagi2019ha}, saliency detection \cite{liu2018picanet, zhang2018progressive}, image super-resolution \cite{zhang2018image}, and so on. Attention mechanisms are becoming one of the most popular approaches for learning some important and useful information.
There are mainly two directions of using the attention, channel-wise attention, and spatial attention. Channel-wise attention \cite{Chen2017SCA} is to learn different weights by multi-channels to emphasize on some special features, while spatial attention \cite{Gregor2015DRAW,Chen2016Attention} is to learn different weights by different spatial information in a map.

Regarding to the diagnosis of COVID-19, there are also some attention-related networks are proposed to highlight the important area for making clinical decisions. For example, Ouyang \emph{et al.} \cite{9095328} designed a dual-sampling attention network, with uniform sampling and size-balanced sampling, to ensure that the network can make decisions based on highlighted infection regions. Fan \emph{et al.} \cite{9098956} utilized edge-attention to enhance the representations. Li et al. \cite{li2020artificial} tried to took the attention heatmap to illustrate the abnormal regions.
Similar to these studies, we also take the attention methods to recognize the important regions which are having effects on decisions. Moreover, to our best knowledge, we are the first to propose using lung-window features as the attention map, while learning most information from the mediastinal window.

\section{Methods}
\label{methods}
\subsection{Dual Window Network}
\begin{figure}[htbp]
    \centerline{\includegraphics[width=150mm]{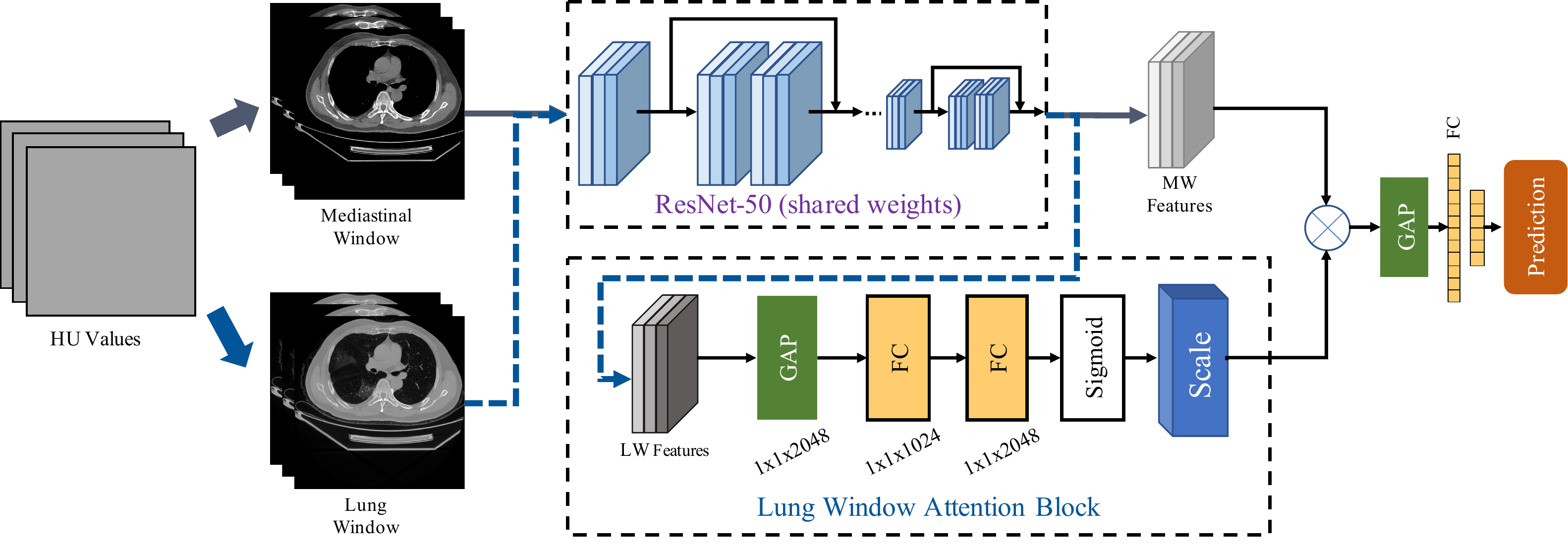}}
    \vspace{-0cm}
    \caption{Architecture of the DWRNet. The input HU values will be transformed into the lung window images and the mediastinal window images with thresholding methods. Both CT windows share the same ResNet-50 and generate corresponding features separately. `GAP' indicates Global Average Pooling layers. FC indicates fully-connected layer. The lung window (LW) features will be fed into the Lung Window Attention Block and rescaled.
    }
    \vspace{-0cm}
    \label{model}
    \end{figure}

The architecture of the DWRNet is shown in Figure.\ref{model}. Before being fed into the framework, the original HU values will be transformed into two CT image window: lung window $X_L$ (HU[-1000, 400]), and mediastinal window $X_M$ (HU[-160, 240]). 
Unlike the conventional two-stream method, our two image windows share the same ResNet-50 model as a visual feature extractor. This design allows us to build a deeper framework with low memory cost. To boost the training speed, we use the ResNet-50 pre-trained on ImageNet.

More specifically, the $X_L$ and $X_M$ are fed into the ResNet separately and get two separate feature vectors $F_L$ and $F_W$:
\begin{equation}
    V_L,V_W=ResNet(X_L,X_M)
\end{equation}
Because the mediastinal window can provide more discriminative features for the diagnosis of severe cases, we treat it as the main CT window.
To enhance the visual features fed into the decision-making process, we use a Lung Window Attention Block (LWA Block), which is invoked by the study \cite{2017Squeeze}, to learn the attention map from the lung window feature channels, and then add this attention information to the mediastinal window features as a guideline:
$FV_M \otimes LWA(V_L)$.

\subsection{Lung Window Attention Block}

The architecture of LWA Block is also shown in Figure \ref{model}. The lung window features and mediastinal window features come from ResNet-50 without fully-connected layers, which have a shape of $8\times8\times2048$. Then we use Global Average Pooling (GAP) \cite{lin2014network} to transform lung window features into $1\times2048$, which will be fed into two fully-connected layers with shapes of $1\times1024$, and $1\times2048$. After two fully-connected layers, the learned weights will be fused with corresponding channels in the mediastinal window features. This block can help to reduce redundant information from the lung window, and highlight the different symptom appearances between two CT image windows.

We fuse attention features directly to the mediastinal window features for two reasons:
1. The lung window it the most common CT window used in clinical practice, which means the lung window can provide more but unnecessary visual features than the mediastinal window. LWA Block can extract more critical features from the massive information provided by the lung window.
2. The mediastinal window is used more often when the lungs have severe symptoms. It means the mediastinal window images contain fewer but discriminative visual features than the lung window. As a result, we apply the attention mechanism to the lung window rather than the mediastinal window.

\subsection{Overall Loss}
After fusing the lung window attention features and the mediastinal window features, the fused information will be fed into a GAP layer and two fully-connected layers. The prediction results will be given after the final Softmax layer:
\begin{equation}
    Pred=Softmax(F(V_M \otimes SE(V_L)))
\end{equation}
The disease course diagnosis is actually a binary classification task, so we use the cross-entropy function as the loss function:
\begin{equation}
    \mathop{\arg\min}_{W} -\frac{1}{Q} \sum_{q=1}^{Q} \frac{1}{N} \sum_{X_n\in\chi}(y_n^q = q) log(Pred(y_n^q=q|X_n;W))
\end{equation}
$\chi=\{X_n\}_{n=1}^N$ denotes the training set, $X_n$ represents $n$-th case of training set. $y^q$ is label vector. In this study, the class labels are used in a back-propagation procedure to update the network weights in the convolutional, and learn the most relevant features in the fully-connected layers. $W$ denotes the trainable parameters of the proposed framework. $Q$ is equal to $2$, which indicates two classes in our study.

\section{Experiments}
\label{experiments}
\subsection{Evaluation Dataset and Experimental Settings}
\wangql{To evaluate the framework proposed in this paper, we analyze 235 adult cases of COVID-19 cases from Hubei province collected by the First Affiliated Hospital of Army Medical University,} which contain more than 161 thousand COVID-19 CT \wangql{slices}. All cases were labeled by the radiologists into two categories: common cases (163) and severe cases (72).
The scale of the dataset is relatively smaller than the binary classification dataset, since acquiring cases with disease course labels costs more efforts. 

Raw data collected from the central hospital may contain more than one series of CT slices, which cannot be learned by deep learning methods directly. To facilitate the training of deep learning models and switch of the CT images window, we only keep the slices with the largest number and the smallest slice-thickness. Then we transform all CT images into HU values to keep unified standards.
Before training, we further transform all HU values into the lung window image and the mediastinal window image.

We randomly select 165 cases (70\%) as the training set, and the rest cases are used as the testing set. The training set and testing set have the same data distribution. The details of our dataset are listed in Table.\ref{dataset}. Due to the small number of severe cases, the cross validation is not adopted.

\begin{table}[htbp]
    \begin{center}
    \begin{tabular}{c|c|c}
    \hline
    \textbf{\textit{Dataset}} & \textbf{\textit{Common Type}}& \textbf{\textit{Severe Type}}\\
    \hline
    All Data & 163 & 72 \\
    Training Set & 114 & 51 \\
    Testing Set & 49 & 21 \\
    \hline
    \end{tabular}
    \vspace{-0cm}
    \end{center}
    \vspace{-0cm}\
    \caption{Details of the dataset. We have 163 common type cases and 72 severe type in total. The training set contains 70\% of the dataset, and the rest will be used as the testing set.}
    \label{dataset}
    \end{table}

Our framework is built on the Tensorflow 1.4 \cite{abadi2016tensorflow}, and run on a Tesla-V100 GPU. The learning rate is set to $5\times 10^{-4}$, and drops 50\% each 800 steps. During each epoch, each case will be trained 12 times, and the whole framework will be trained 12 epochs.

\subsection{Experimental Analyse}
In this section, we will discuss the experimental performances of the proposed framework.
We build five models: Recurrent CNN with original HU images (RCNN-HU), Recurrent CNN with lung window images (RCNN-LW), Recurrent CNN with mediastinal window images (RCNN-MW), DWRNet with mediastinal window attention (DWRNet-M), and DWRNet with lung window attention (DWRNet-L), which is final design of our framework. It should be noted that DWRNet-M uses lung window images as its main image window, and on the contrary, DWRNet-L uses mediastinal window images as its main image window.

Experimental results are listed in Table.\ref{results}.
We test all models five times and calculate the average sensitivity, specificity, and accuracy for comprehensive comparisons. The sensitivity indicates the ability to classify positive severe COVID-19 cases; the specificity indicates the ability to classify negative common COVID-19 cases. 

\begin{table}[htbp]
    \begin{center}
    \begin{tabular}{c|c|c|c|c|c|c|c}
    \hline
    \textbf{\textit{Data}}  & \textbf{\textit{EVA}} & \textbf{\textit{Test1}} & \textbf{\textit{Test2}} & \textbf{\textit{Test3}} & \textbf{\textit{Test4}}& \textbf{\textit{Test5}}& \textbf{\textit{Ave}}\\
    \hline
    \multirow{3}{*}{RCNN-HU} & SEN & 0.5714 & 0.5714 & 0.6190 & 0.5714 & 0.5714 & 0.5810 \\
    &SPE& 0.9388&0.9796&0.9796&0.9592&0.9592&0.9633\\
    &ACC&0.8286&0.8571&0.8714&0.8429&0.8429&0.8486 \\
    \hline
    \multirow{3}{*}{RCNN-LW} & SEN&0.5714&0.5714&0.6667&0.6667&0.6667&0.6667 \\
    &SPE&0.9184&0.9592&0.9592&0.9592&0.9184&0.9429 \\
    &ACC&0.8143&0.8429&0.8714&0.8714&0.8429&0.8486 \\
    \hline
    \multirow{3}{*}{RCNN-MW} & SEN&0.7143&0.6667&0.7619&0.6667&0.6667&0.6952  \\
    &SPE&0.9592&0.9592&0.9796&0.9388&0.9796&0.9633 \\
    &ACC&0.8857&0.8714&0.9143&0.8571&0.8857&0.8829 \\
    \hline
    \multirow{3}{*}{DWRNet-M} & SEN&0.6667&0.6667&0.6667&0.7143&0.6667&0.6762\\
    &SPE&0.9592&0.9388&0.9592&0.9592&0.9388&0.9510 \\
    &ACC&0.8714&0.8714&0.8714&0.8857&0.8571&0.8714 \\
    \hline
    \multirow{3}{*}{DWRNet-L} & SEN&0.7619&0.7143&0.7619&0.7143&0.7143&\textbf{0.7333} \\
    &SPE&0.9796&0.9796&0.9796&0.9796&0.9796&\textbf{0.9796} \\
    &ACC&0.9143&0.9000&0.9143&0.9000&0.9000&\textbf{0.9057}\\
    \hline
    \end{tabular}
    \vspace{-0cm}
    \end{center}
    \vspace{-0cm}
    \caption{Experimental Results. Five models are tested: Recurrent CNN with original HU images (RCNN-HU), Recurrent CNN with lung window images (RCNN-LW), Recurrent CNN with mediastinal window images (RCNN-MW), DWRNet with mediastinal window attention (DWRNet-M), and DWRNet with lung window attention (DWRNet-L). ACC, SPE and SEN indicate accuracy, specificity, and sensitivity.}
    \label{results}
    \end{table}

We can see that RCNN-HU and RCNN-LW have similar accuracy, but RCNN-LW has higher sensitivity. RCNN-MW has an accuracy of 88.29\%, which is higher than RCNN-HU and RCNN-LW, it means the mediastinal window is more suitable for the classification of disease courses. Our DWRNet-L has the highest performances in all accuracy, sensitivity, and specificity. DWRNet-M has a lower accuracy than RCNN-MW, but higher than RCNN-LW.

\subsubsection{Effect of CT Image Window Settings}
In this section, we discuss the influences of CT image windows. The baseline Recurrent CNN is trained with HU images, which are directly sampled from CT.
The sensitivity, specificity, and accuracy of the baseline are 58.10\%, 96.33\%, and 84.86\%.
As can be seen, this model tends to predict cases as common cases. When the CT image is adjusted to the lung window, the performance of sensitivity improves by 8.57\%. However, the accuracy of RCNN-LW is still 84.86\%.
When we use the mediastinal window as inputs, the sensitivity is improved to 69.52\%, and the accuracy is improved to 88.29\%.

It is a very interesting phenomenon since most of the existing studies mainly focus on analyzing CT with the lung window. According to our experiments, the mediastinal window is more suitable for classifying common/severe cases. We believe it is because severe COVID-19 cases contain more severe symptoms that may lead to density changes, such as pulmonary consolidation, and the mediastinal window has its unique advantage in observing soft tissues. As a result, we use the mediastinal window as the main image window in our DWRNet.

\subsubsection{Effect of Different Main Image Window}
\label{exp2}
In this section, we will compare two versions of DWRNet: DWRNet-M, DWRNet-L. DWRNet-M takes the lung window as its main image window and uses the mediastinal window to guide the framework using the attention mechanism. On the contrary, DWRNet-L takes the mediastinal window as its main image window and uses the lung window as its guideline.

As can be seen in the Table~\ref{results}, DWRNet-M achieves higher accuracy than RCNN-LW, but lower than RCNN-MW.
DWRNet-L achieves the highest sensitivity, specificity, and accuracy among all models, which reflect the effect of the mediastinal window and the attention map from the lung window.

Moreover, the proposed DWRNet also shows better stability. We calculate the standard deviations of each model in five tests, which are listed in Table.\ref{sd}. As can be seen, both versions of DWRNets have better stability than RCNN with a single image window. DWRNet-M has the lowest standard deviation in specificity, and DWRNet-L has the lowest standard deviations in both specificity and accuracy. It means the attention information of additional CT image windows can improve the performances of the proposed framework and improve the stability.

According to our experimental results, a few conclusions can be summarized:
1. The mediastinal window is more suitable for the diagnosis of COVID-19 disease course since it can provide more discriminative visual features for severe cases.
2. The additional attention map provided by LWA Block can improve the performances and stability of the framework.

\begin{table}[htbp]
    \begin{center}
    \begin{tabular}{c|c|c}
    \hline
    \textbf{\textit{Model}} & \textbf{\textit{EVA}} & \textbf{\textit{Standard Deviation}}\\
    \hline
    \multirow{3}{*}{RCNN-LW} & SEN&0.0494 \\
    &SPE&0.0205 \\
    &ACC&0.0220 \\
    \hline
    \multirow{3}{*}{RCNN-MW} & SEN&0.0399  \\
    &SPE&0.0156 \\
    &ACC&0.0194 \\
    \hline
    \multirow{3}{*}{DWRNet-M} & SEN&\textbf{0.0195}\\
    &SPE&0.0101 \\
    &ACC&0.0091 \\
    \hline
    \multirow{3}{*}{DWRNet-L} & SEN&0.0240\\
    &SPE&\textbf{0.0}\\
    &ACC&\textbf{0.0071}\\
    \hline
    \end{tabular}
    \vspace{-0cm}
    \end{center}
    \vspace{-0cm}
    \caption{Standard Deviations of Sensitivity, Specificity, and Accuracy. Four models are tested: Recurrent CNN with lung window images (RCNN-LW), Recurrent CNN with mediastinal window images (RCNN-MW), DWRNet with mediastinal window attention (DWRNet-M), and DWRNet with lung window attention (DWRNet-L). }
    \label{sd}
    \end{table}

\subsubsection{Effectiveness of LWA Block}
\begin{figure}[htbp]
    \centerline{\includegraphics[width=110mm]{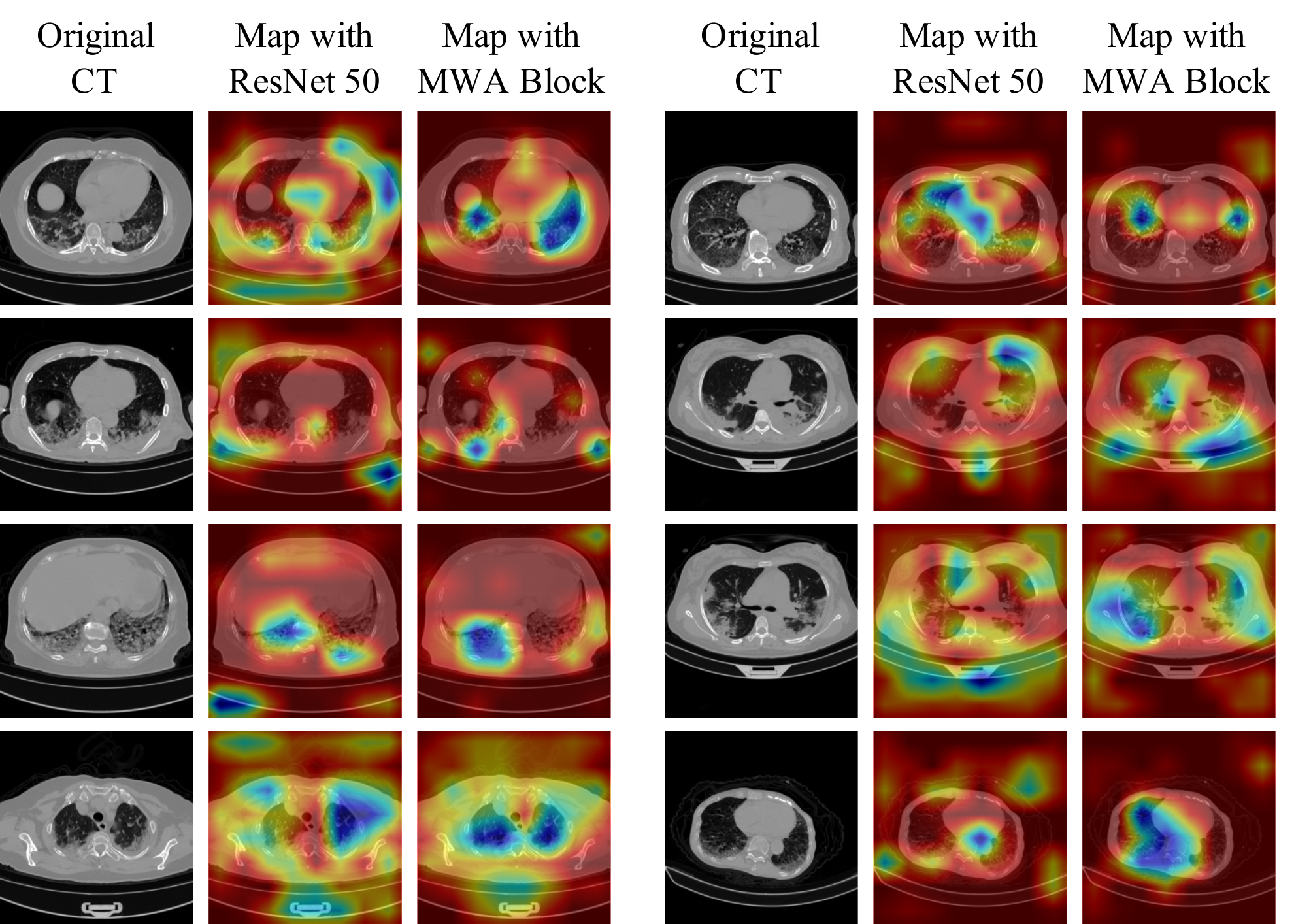}}
    \vspace{-0cm}
    \caption{Feature Maps of DWRNet-M. Feature Maps of DWRNet-M. \wangql{This framework uses lung window as its main image window, and the MWA (Mediastinal-Window Attention) Block learns attention features from the mediastinal window. Eight slices of COVID-19 CT are shown. For each slice, we show the original CT, activation maps of ResNet, and activation maps of our framework.}
    }
    \vspace{-0cm}
    \label{map1}
    \end{figure}

\begin{figure}[htbp]
    \centerline{\includegraphics[width=110mm]{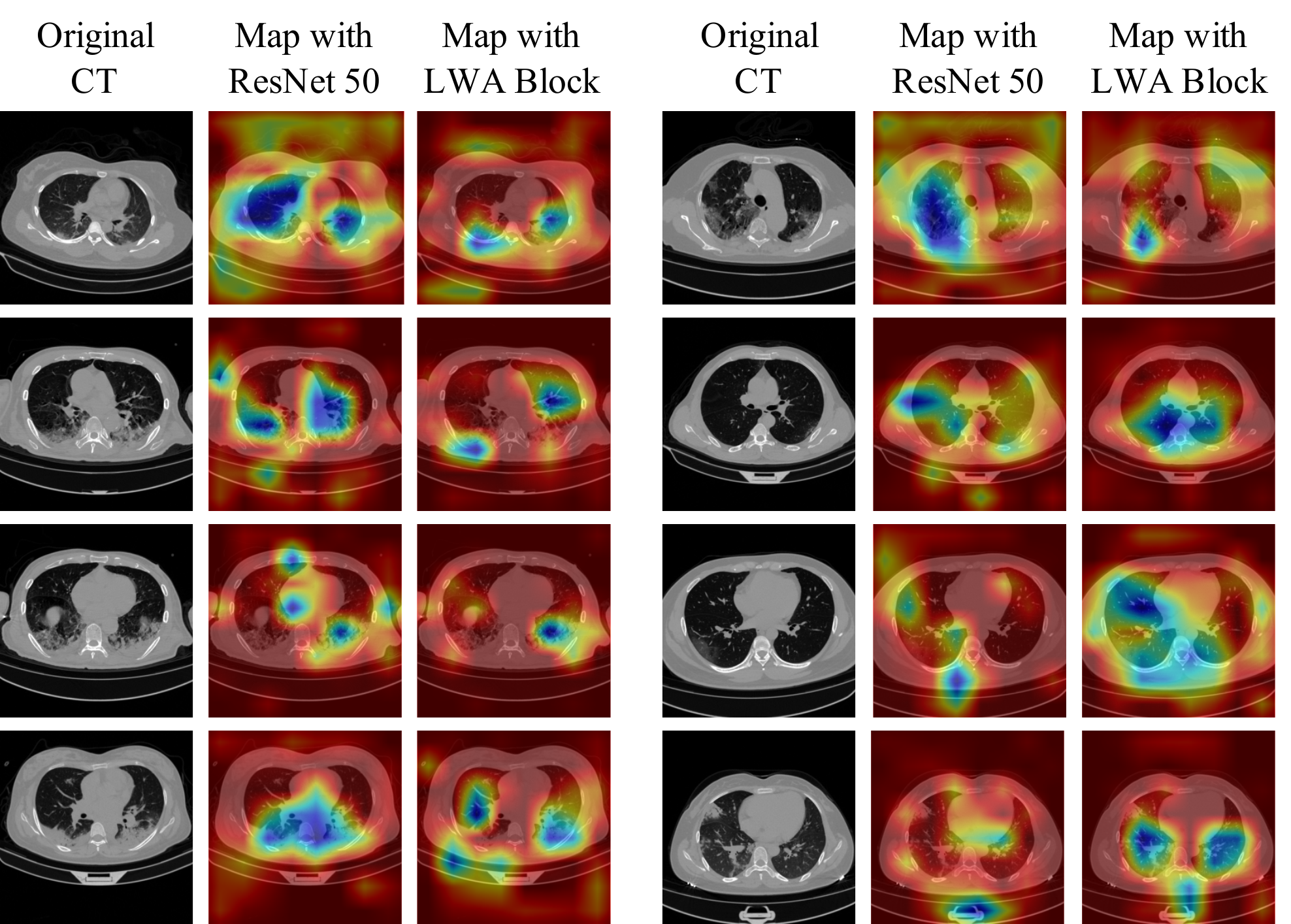}}
    \vspace{-0cm}
    \caption{Feature Maps of DWRNet-L.\wangql{This framework uses mediastinal window as its main image window, and the LWA Block learns attention features from the lung window.}
    }
    \vspace{-0cm}
    \label{map2}
    \end{figure}

To further demonstrate the effectiveness of the LWA Block, we show activation maps of DWRNet-M and DWRNet-L in Figure \ref{map1} and Figure \ref{map2}.
As shown in both figures, the activation maps with LWA Block concentrate more on the areas of lungs, which means the LWA Block with additional CT windows can help the DWRNet focus on the important areas as our design.
In Section~\ref{exp2}, we observe that the DWRNet-L has better performance than DWRNet-W, which can also be explained by Figure~\ref{map2}: the mediastinal window images can help the framework better focus on the areas of the lungs, which can significantly provide better receptive fields for neural networks. Moreover, the DWRNet-L can correct the focus with the help of LWA Block and achieve better and more stable performances.

\section{Discussion}
\label{discussion}
As can be seen, the highest sensitivity is still lower than 80\%, which may limit the clinical practice of DWRNet. We look into the \emph{Diagnosis and Treatment Protocol for COVID-19 Patients} from \emph{China National Health Commission} and find some facts that can explain why the sensitivity is low.

According to the \emph{Diagnosis and Treatment Protocol for COVID-19 Patients} \cite{china-protocol}, the diagnosis of severe cases need to consider four facts: 
1) Whether the respiration rate is higher than 30 times per minute; 
2) Whether the oxygen saturation is lower than 93\% at rest; 
3) Arterial oxygen partial pressure (PaO2)/Inspired oxygen concentration (FiO2) is lower than 300 mmHg; and 
4) Progressive aggravation of clinical symptoms observing lung images. 
In other words, CT is not the only discriminative information that can be used for classifying common/severe cases. The only solution to improving the sensitivity or detecting severe cases is to fuse multiple information resources, such as respiration rate, oxygen saturation, et al.
However, collecting such information is a time-consuming task, and the deep learning framework needs to be updated for learning more information. We are collecting more data and our future work will focus on this task.

\section{Conclusions}
\label{conclusions}
This paper proposes a novel Dual Window RCNN Network (DWRNet), which treats the mediastinal window as the main visual information resources, and uses attenuation map from the lung window to improve the overall performances.
\wangql{In this paper, instead of picking up specific slices for deep learning framework, CT slices are treated as videos and analyzed by RCNN, which can keep 3D information and reduce calculation burden at the same time.}
A Lung Window Attention (LWA Block) is introduced to fuse lung window features into mediastinal window feature as the guidance.
Our experiments show that the mediastinal window is more suitable for the diagnosis of disease course.
Moreover, the fused and representative features can improve the stability of the DWRNet, and improve the accuracy by 4\% compared to the baseline.

\section{Acknowledgments}
This work was partially supported by the National Key R\&D Project of China (Grant No. 2018YFB2101200), the National Natural Science Foundation of China (Grant No. 61772093), the Special General Project of Chongqing Technology Innovation and Application Development (Grant No. cstc2019jscx-msxmX0104), and the Medical Science and Technology Youth Training Program (Grant No. 20QNPY013).

\bibliography{mybibfile}

\begin{thebibliography}{10}
\expandafter\ifx\csname url\endcsname\relax
  \def\url#1{\texttt{#1}}\fi
\expandafter\ifx\csname urlprefix\endcsname\relax\def\urlprefix{URL }\fi
\expandafter\ifx\csname href\endcsname\relax
  \def\href#1#2{#2} \def\path#1{#1}\fi

\bibitem{article}
L.~Dong, J.~Tian, S.~He, C.~Zhu, J.~Wang, C.~Liu, J.~Yang, Possible vertical
  transmission of sars-cov-2 from an infected mother to her newborn, JAMA The
  Journal of the American Medical Association 323.
\newblock \href {http://dx.doi.org/10.1001/jama.2020.4621}
  {\path{doi:10.1001/jama.2020.4621}}.

\bibitem{who-2020cases}
W.~H.~O. (WHO), Weekly operational update on covid-19,
  \url{https://www.who.int/emergencies/diseases/novel-coronavirus-2019/situation-reports}.

\bibitem{9090149}
Y.~{Oh}, S.~{Park}, J.~C. {Ye}, Deep learning covid-19 features on cxr using
  limited training data sets, IEEE Transactions on Medical Imaging 39~(8)
  (2020) 2688--2700.

\bibitem{2020Prior}
J.~Wang, Y.~Bao, Y.~Wen, H.~Lu, D.~Qian, Prior-attention residual learning for
  more discriminative covid-19 screening in ct images, IEEE Transactions on
  Medical Imaging PP~(99) (2020) 1--1.

\bibitem{WANG2020}
S.-H. Wang, V.~V. Govindaraj, J.~M. Górriz, X.~Zhang, Y.-D. Zhang, Covid-19
  classification by fgcnet with deep feature fusion from graph convolutional
  network and convolutional neural network, Information Fusion\href
  {http://dx.doi.org/https://doi.org/10.1016/j.inffus.2020.10.004}
  {\path{doi:https://doi.org/10.1016/j.inffus.2020.10.004}}.

\bibitem{china-protocol}
N.~H.~C. of~the People's Republic~of China, Diagnosis and treatment protocol
  for covid-19 patients (tentative 8th edition),
  \url{http://en.nhc.gov.cn/2020-09/07/c_81565.htm}.

\bibitem{macmahon2017guidelines}
H.~Macmahon, D.~P. Naidich, J.~M. Goo, K.~S. Lee, A.~N. Leung, J.~R. Mayo,
  A.~C. Mehta, Y.~Ohno, C.~A. Powell, M.~Prokop, et~al., Guidelines for
  management of incidental pulmonary nodules detected on ct images: From the
  fleischner society 2017, Radiology 284~(1) (2017) 228--243.

\bibitem{2018Effect}
H.~Ahn, K.~W. Lee, K.~H. Lee, J.~Kim, K.~Kim, J.~H. Chung, C.~T. Lee, Effect of
  computed tomography window settings and reconstruction plane on 8th edition
  t-stage classification in patients with lung adenocarcinoma manifesting as a
  subsolid nodule, European Journal of Radiology 98 (2018) 130.

\bibitem{2018Radiologic}
M.~Yanagawa, M.~Kusumoto, T.~Johkoh, M.~Noguchi, N.~Tomiyama,
  Radiologic-pathologic correlation of solid portions on thin-section ct images
  in lung adenocarcinoma: A multicenter study, Clinical Lung Cancer 19~(3).

\bibitem{Yao2016Value}
Yao, Gang, Value of window technique in diagnosis of the ground glass opacities
  in patients with non-small cell pulmonary cancer, Oncology Letters 12~(5)
  (2016) 3933--3935.

\bibitem{Upchurch2017Community}
C.~P. Upchurch, C.~G. Grijalva, R.~G. Wunderink, D.~J. Williams, G.~W. Waterer,
  E.~J. Anderson, Y.~Zhu, E.~M. Hart, F.~Carroll, A.~M. Bramley,
  Community-acquired pneumonia visualized on ct scans but not chest
  radiographs: Pathogens, severity, and clinical outcomes, Chest 153~(3).

\bibitem{XIE2018102}
Y.~Xie, J.~Zhang, Y.~Xia, M.~Fulham, Y.~Zhang, Fusing texture, shape and deep
  model-learned information at decision level for automated classification of
  lung nodules on chest ct, Information Fusion 42 (2018) 102 -- 110.
\newblock \href
  {http://dx.doi.org/https://doi.org/10.1016/j.inffus.2017.10.005}
  {\path{doi:https://doi.org/10.1016/j.inffus.2017.10.005}}.

\bibitem{shin2016deep}
H.-C. Shin, H.~R. Roth, M.~Gao, L.~Lu, Z.~Xu, I.~Nogues, J.~Yao, D.~Mollura,
  R.~M. Summers, Deep convolutional neural networks for computer-aided
  detection: Cnn architectures, dataset characteristics and transfer learning,
  IEEE Transactions on Medical Imaging 35~(5) (2016) 1285--1298.

\bibitem{gao2018holistic}
M.~Gao, U.~Bagci, L.~Lu, A.~Wu, M.~Buty, H.-C. Shin, H.~Roth, G.~Z. Papadakis,
  A.~Depeursinge, R.~M. Summers, et~al., Holistic classification of ct
  attenuation patterns for interstitial lung diseases via deep convolutional
  neural networks, Computer Methods in Biomechanics and Biomedical Engineering:
  Imaging \& Visualization 6~(1) (2018) 1--6.

\bibitem{9078812}
Q.~{Wang}, D.~{Yang}, Z.~{Li}, X.~{Zhang}, C.~{Liu}, Deep regression via
  multi-channel multi-modal learning for pneumonia screening, IEEE Access 8
  (2020) 78530--78541.

\bibitem{zu2020coronavirus}
Z.~Y. Zu, Jiang, P.~P. Xu, W.~Chen, Q.~Q. Ni, G.~Lu, L.~J. Zhang, Coronavirus
  disease 2019 (covid-19): A perspective from china, Radiology (2020)
  200490--200490.

\bibitem{9086482}
H.~{Kang}, L.~{Xia}, F.~{Yan}, Z.~{Wan}, F.~{Shi}, H.~{Yuan}, H.~{Jiang},
  D.~{Wu}, H.~{Sui}, C.~{Zhang}, D.~{Shen}, Diagnosis of coronavirus disease
  2019 (covid-19) with structured latent multi-view representation learning,
  IEEE Transactions on Medical Imaging 39~(8) (2020) 2606--2614.

\bibitem{9098956}
D.~{Fan}, T.~{Zhou}, G.~{Ji}, Y.~{Zhou}, G.~{Chen}, H.~{Fu}, J.~{Shen},
  L.~{Shao}, Inf-net: Automatic covid-19 lung infection segmentation from ct
  images, IEEE Transactions on Medical Imaging 39~(8) (2020) 2626--2637.

\bibitem{9097297}
X.~{Wang}, X.~{Deng}, Q.~{Fu}, Q.~{Zhou}, J.~{Feng}, H.~{Ma}, W.~{Liu},
  C.~{Zheng}, A weakly-supervised framework for covid-19 classification and
  lesion localization from chest ct, IEEE Transactions on Medical Imaging
  39~(8) (2020) 2615--2625.

\bibitem{simonyan2014two-stream}
K.~Simonyan, A.~Zisserman, Two-stream convolutional networks for action
  recognition in videos (2014).
\newblock \href {http://arxiv.org/abs/1406.2199} {\path{arXiv:1406.2199}}.

\bibitem{he2016deep}
K.~He, X.~Zhang, S.~Ren, J.~Sun, Deep residual learning for image recognition
  (2015).
\newblock \href {http://arxiv.org/abs/1512.03385} {\path{arXiv:1512.03385}}.

\bibitem{Guo_2020_CVPR}
L.~Guo, J.~Liu, X.~Zhu, P.~Yao, S.~Lu, H.~Lu, Normalized and geometry-aware
  self-attention network for image captioning, in: IEEE/CVF Conference on
  Computer Vision and Pattern Recognition (CVPR), 2020.

\bibitem{you2016image}
Q.~You, H.~Jin, Z.~Wang, C.~Fang, J.~Luo, Image captioning with semantic
  attention, in: IEEE/CVF Conference on Computer Vision and Pattern Recognition
  (CVPR), 2016, pp. 4651--4659.

\bibitem{Chen2017SCA}
L.~Chen, H.~Zhang, J.~Xiao, L.~Nie, J.~Shao, W.~Liu, T.~S. Chua, Sca-cnn:
  Spatial and channel-wise attention in convolutional networks for image
  captioning, in: IEEE/CVF Conference on Computer Vision and Pattern
  Recognition (CVPR), 2017.

\bibitem{Yu2018Learning}
C.~Yu, J.~Wang, C.~Peng, C.~Gao, G.~Yu, N.~Sang, Learning a discriminative
  feature network for semantic segmentation, in: IEEE/CVF Conference on
  Computer Vision and Pattern Recognition (CVPR), 2018.

\bibitem{Choi_2020_CVPR}
S.~Choi, J.~T. Kim, J.~Choo, Cars can't fly up in the sky: Improving
  urban-scene segmentation via height-driven attention networks, in: IEEE/CVF
  Conference on Computer Vision and Pattern Recognition (CVPR), 2020.

\bibitem{Jiang_2020_CVPR}
X.~Jiang, L.~Zhang, M.~Xu, T.~Zhang, P.~Lv, B.~Zhou, X.~Yang, Y.~Pang,
  Attention scaling for crowd counting, in: IEEE/CVF Conference on Computer
  Vision and Pattern Recognition (CVPR), 2020.

\bibitem{sindagi2019ha}
V.~A. Sindagi, V.~M. Patel, Ha-ccn: Hierarchical attention-based crowd counting
  network, IEEE Transactions on Image Processing 29 (2019) 323--335.

\bibitem{liu2018picanet}
N.~Liu, J.~Han, M.-H. Yang, Picanet: Learning pixel-wise contextual attention
  for saliency detection, in: IEEE/CVF Conference on Computer Vision and
  Pattern Recognition (CVPR), 2018, pp. 3089--3098.

\bibitem{zhang2018progressive}
X.~Zhang, T.~Wang, J.~Qi, H.~Lu, G.~Wang, Progressive attention guided
  recurrent network for salient object detection, in: IEEE/CVF Conference on
  Computer Vision and Pattern Recognition (CVPR), 2018, pp. 714--722.

\bibitem{zhang2018image}
Y.~Zhang, K.~Li, K.~Li, L.~Wang, B.~Zhong, Y.~Fu, Image super-resolution using
  very deep residual channel attention networks, in: Proceedings of the
  European Conference on Computer Vision (ECCV), 2018, pp. 286--301.

\bibitem{Gregor2015DRAW}
K.~Gregor, I.~Danihelka, A.~Graves, D.~J. Rezende, D.~Wierstra, Draw: a
  recurrent neural network for image generation, in: International Conference
  on Machine Learning (ICML), 2015, pp. 1462--1471.

\bibitem{Chen2016Attention}
L.~C. Chen, Y.~Yang, J.~Wang, W.~Xu, A.~Yuille, Attention to scale: Scale-aware
  semantic image segmentation, in: IEEE/CVF Conference on Computer Vision and
  Pattern Recognition (CVPR), 2016, pp. 3640--3649.

\bibitem{9095328}
X.~{Ouyang}, J.~{Huo}, L.~{Xia}, F.~{Shan}, J.~{Liu}, Z.~{Mo}, F.~{Yan},
  Z.~{Ding}, Q.~{Yang}, B.~{Song}, F.~{Shi}, H.~{Yuan}, Y.~{Wei}, X.~{Cao},
  Y.~{Gao}, D.~{Wu}, Q.~{Wang}, D.~{Shen}, Dual-sampling attention network for
  diagnosis of covid-19 from community acquired pneumonia, IEEE Transactions on
  Medical Imaging 39~(8) (2020) 2595--2605.

\bibitem{li2020artificial}
L.~Li, L.~Qin, Z.~Xu, Y.~Yin, X.~Wang, B.~Kong, J.~Bai, Y.~Lu, Z.~Fang,
  Q.~Song, et~al., Artificial intelligence distinguishes covid-19 from
  community acquired pneumonia on chest ct, Radiology.

\bibitem{2017Squeeze}
J.~Hu, L.~Shen, S.~Albanie, G.~Sun, E.~Wu, Squeeze-and-excitation networks,
  IEEE Transactions on Pattern Analysis and Machine Intelligence PP~(99).

\bibitem{lin2014network}
M.~Lin, Q.~Chen, S.~Yan, Network in network (2014).
\newblock \href {http://arxiv.org/abs/1312.4400} {\path{arXiv:1312.4400}}.

\bibitem{abadi2016tensorflow}
M.~Abadi, A.~Agarwal, P.~Barham, E.~Brevdo, Z.~Chen, C.~Citro, et~al.,
  Tensorflow: Large-scale machine learning on heterogeneous distributed systems
  (2016).
\newblock \href {http://arxiv.org/abs/1603.04467} {\path{arXiv:1603.04467}}.

\end{thebibliography}

\end{document}